\documentclass[fleqn,usenatbib]{mnras}

\usepackage{newtxtext,newtxmath}
\usepackage[T1]{fontenc}

\DeclareRobustCommand{\VAN}[3]{#2}
\let    \VANthebibliography\thebibliography
\def\thebibliography{\DeclareRobustCommand{\VAN}[3]{##3}\VANthebibliography}

\usepackage{graphicx}	% Including figure files
\usepackage{amsmath}	% Advanced maths commands
\usepackage{xfrac}
\usepackage{bm}
\usepackage{orcidlink}
\title[Depolarization and Faraday effects in AGN Jets]{Depolarization and Faraday effects in AGN Jets}

\author[E. Yushkov et al.]{
E. Yushkov$^{1, 2}$\orcidlink{0000-0002-9810-1541},
I.N. Pashchenko$^{3}$\orcidlink{0000-0002-9404-7023},
D. Sokoloff,$^{1,4}$\thanks{E-mail: sokoloff.dd@gmail.com}\orcidlink{0000-0002-3441-0863}
and G. Chumarin$^{1}$
\\
$^{1}$Moscow State University, Leninskie Gory 1, Moscow, Russia\\
$^{2}$Space Research Institute of the Russian Academy of Sciences (IKI), Profsoyuznaya st., 84/32, Moscow, Russia\\
$^{3}$Lebedev Physical Institute of the Russian Academy of Sciences, Leninsky prospekt 53, 119991 Moscow, Russia\\
$^{4}$IZMIRAN, Kaluzhskoe Shosse 4, Troitsk, Moscow, Russia}

\date{Accepted XXX. Received YYY; in original form ZZZ}
\pubyear{2024}

\begin{document}
\label{firstpage}
\pagerange{\pageref{firstpage}--\pageref{lastpage}}
\maketitle

\begin{abstract}
Radio interferometric observations of Active Galactic Nuclei (AGN) jets reveal the significant linear polarization of their synchrotron radiation that changes with frequency due to the Faraday rotation. It is generally assumed that such depolarization could be a powerful tool for studying the magnetized plasma in the vicinity of the jet. However, depolarization could also occur within the jet if the emitting and rotating plasma are co-spatial (i.e. the internal Faraday rotation). Burn obtained very simple dependence of the polarization on the wavelength squared for the discrete source and resolved slab that is widely used for interpreting the depolarization of AGN jets. However it ignores the influence of the non-uniform large scale magnetic field of the jet on the depolarization. Under the simple assumptions about the possible jet magnetic field structures we obtain the corresponding generalizations of Burn's relation widely used for galaxies analysis. We show that the frequency dependencies of the Faraday rotation measure and polarization angle in some cases allow to estimate the structures of the jets magnetic fields.
\end{abstract}

\begin{keywords}
galaxies: active, galaxies: jets, Physical data and processes: radiation mechanisms: non-thermal
\end{keywords}

%%%%%%%%%%%%%%%%% BODY OF PAPER %%%%%%%%%%%%%%%%%%
\section{Introduction}
\label{sec:introduction}

% \textbf{Rename the section to Motivation?}

Large scale magnetic fields are thought to play \textcolor{black}{a crucial role} in launching, acceleration and collimation \textcolor{black}{of jets in active galactic nuclei} \citep[][]{2001Sci...291...84M, 2006MNRAS.368.1561M,2007MNRAS.380...51K,2019ARA&A..57..467B}. Observationally, the most detailed information about the magnetic fields in AGN jets comes \textcolor{black}{from polarimetric} VLBI experiments \cite[e.g.,][]{gabuzda2018magnetic}.
% The synchrotron radiation analysis of relativistic electrons is based on one of the typical features of this radiation -- {significant linear} polarization.
\textcolor{black}{The radiation of parsec-scale AGN jets in the radio band, which is a synchrotron radiation of the ultra relativistic electrons in the jet magnetic field, is significantly linearly polarized \citep{2023MNRAS.520.6053P}.}
The behavior of synchrotron radiation, passing through the magnetically active media, which changes its {linear} polarization {in a frequency dependent manner (i.e. Faraday effect)}, opens up an indirect possibility of reconstructing {magnetic} fields based on the study of such depolarization.

\textcolor{black}{While the ultra relativistic electrons are capable for the emission of the synchrotron radiation, the Faraday effects are mainly driven by the thermal magnetized plasma or, in general, the low energy electrons \citep{1977ApJ...214..522J}. These two populations of particles could be co-spatial, producing the internal depolarization \citep{burn1966depolarization}.} However, the relative importance of the external and internal \textcolor{black}{Faraday effects} is still not clear. 

\textcolor{black}{The observed polarization patterns} of AGN jets suggest the presence of the large-scale helical magnetic field, see \cite{2018Galax...7....5G} for a review.
%\citep{2000MNRAS.319.1109G,2005MNRAS.356..859P,2005MNRAS.360..869L,2013MNRAS.436.3341Z,2014MNRAS.444..172G,2023MNRAS.523.3615Z,2023MNRAS.520.6053P}, 
Multifrequency polarimetric observations reveal
% provide the clues on the particle population that is unobserved directly via the Faraday effects \citep{1998Natur.395..457W,2009ApJ...696..328H,2011AAS...21714234W,2018Galax...6...17H}, as well as
the large scale magnetic field component also in and around the jets \textcolor{black}{on kpc and pc scales} \citep{1984ApJS...54..291P,2002PASJ...54L..39A,2011ApJ...733...11G,2015A&A...583A..96G,2016Galax...4...18M,2016ApJ...825...59A,2017Galax...5...61K,2017MNRAS.472.1792G,2021ApJ...923L...5P} by the transverse gradients of the Faraday Rotation Measure, $RM$. Both observational data \citep{2021ApJ...910...35L} and GRMHD simulations \citep{2010ApJ...725..750B} suggest that such gradients are due to the magnetized sheath around the radiating jet plasma, while there is the evidence that the observed transverse gradients could trace the inner structure of the magnetic field \citep[e.g. in ``magnetic tower'' model,][]{2009MNRAS.400....2M,2013MNRAS.431..695M,2016A&A...591A..61C,2017Galax...5...11G,2018A&A...612A..67G}. Moreover, ``inverse`` or ``anomalous`` depolarization observed in some sources \citep{2012AJ....144..105H} with transverse $RM$ gradient could imply the internal Faraday rotation in a helical magnetic field \citep{sokoloff1998depolarization,2012ApJ...747L..24H}.
The most mass multifrequency polarization VLBI observations of 191 sources from MOJAVE sample at 8, 12 and 15 GHz \citep{2012AJ....144..105H} revealed the depolarization patterns that suggests the external Faraday screen, while some sources require the internal rotation. Detailed 10-frequency (from 1.4 to 15.4 GHz) polarimetric VLBI observations of 20 AGN jets \citep{2017MNRAS.467...83K} are consistent with this results. Broadband (112 subbands over 14 GHz frequency range from 4 to 18 GHz) full-polarization JVLA observations of M87 radio jet {which employed the $QU$-fitting method} revealed the internal Faraday rotation in a helical magnetic field of the jet \citep{2021ApJ...923L...5P}, that is consistent with earlier studies \citep{2011MNRAS.416L.109C}.

The multifrequency polarization data is usually interpreted using various relations between the magnetic field and polarization \citep[e.g.,][]{burn1966depolarization,1991MNRAS.250..726T,sokoloff1998depolarization,2008A&A...487..865R,2009A&A...502...61M}.
{When simultaneous broad-band multi-channel data is available $RM$ synthesis \citep{2005A&A...441.1217B} and $QU$-fitting \citep{2011AJ....141..191F,2012MNRAS.421.3300O} techniques could be employed to, correspondingly, obtain the polarization signal and fit it with a simple analytical models \citep[see][for a review]{2021Galax...9...56P}. While they are shown to be capable to explore the magnetic field in kp-scale jet of M87 \citep{2021ApJ...923L...5P}, spectrally resolve the polarized components of the unresolved radio sources \citep{2018A&A...613A..74P} or to select a jets with a possible transverse $RM$ gradients  \citep{2016ApJ...825...59A}, such dense frequency coverage is not attainable with modern VLBI. Thus, multifrequency VLBI polarization data of parsec-scale AGN jets is analysed using simple depolarization relations \citep{2012AJ....144..105H,2017MNRAS.467...83K} and, when the jet is transversely resolved, the qualitative large scale magnetic field models are employed \citep[][e.g. the toroidal magnetic field to explain the transverse $RM$ gradients]{2002PASJ...54L..39A}.}

One of the quite simple, productive and correspondingly most frequently used {depolarization} relation obtained by \cite{burn1966depolarization} in context of discrete sources was later adopted by \cite{sokoloff1998depolarization} for the discs of the spiral galaxies. Burn's relation is successfully exploited for interpreting the AGN jets internal depolarization \citep[e.g.][]{2012AJ....144..105H,2017MNRAS.467...83K,2019ApJ...871..257P,2021ApJ...923L...5P,2023MNRAS.523.3615Z}. However the geometry of the jets and their expected magnetic configuration are obviously something different from the galactic discs. Of course, one can hope that the differences between discs and jets are not crucial enough to obtain the reasonable estimate of the magnetic field strength, bearing in mind that \cite{sokoloff1998depolarization} provides the examples of the magnetic configurations where the conventional relations lead to the erroneous conclusions. A validation of that approach comes in particular from the fact that \cite{burn1966depolarization} and \cite{sokoloff1998depolarization} departing from different points (discreet sources and discs) arrives to the same scaling. So it looks reasonable to analyse at some point the difference between disc and jets in context of the depolarisation studies. This is just the aim of this paper. To be specific, we consider the depolarization by the large-scale jet magnetic field and intrinsic Faraday rotation {from jet-like cylindrical objects with axial symmetry}.

%%%%%%%%%%%%%%%%%%%%%%%%%%%%%%%%%%%%%%%%%%%%%%%%%
\section{Burn's relation}
\label{sec:burns_relation}

\begin{figure*}
\begin{center}
\includegraphics[width=17cm]{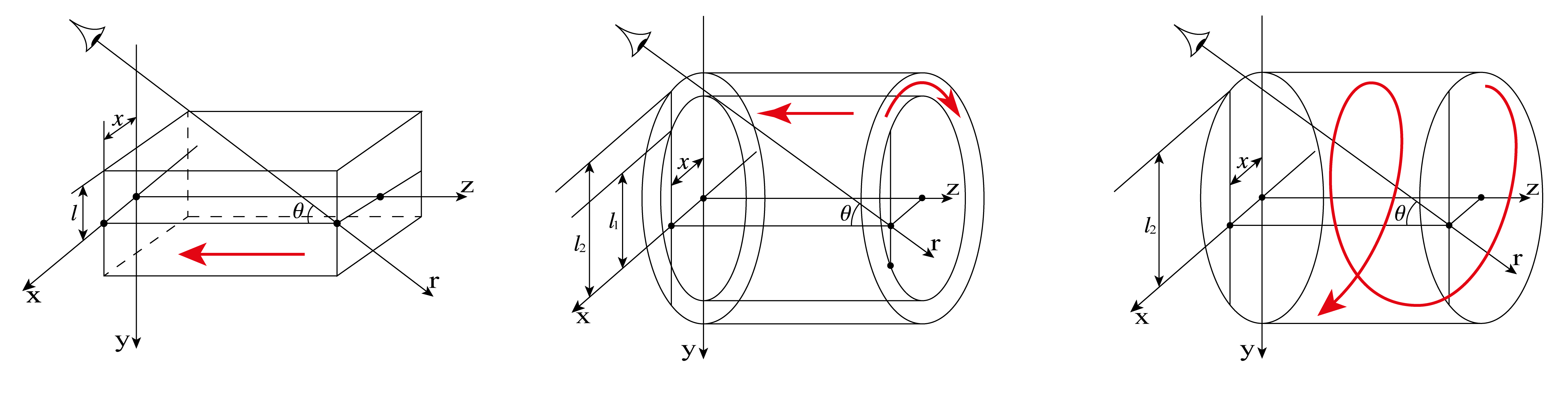}
\caption{Schematic representation of the planar galaxy model (left panel), the two-zone jet model (middle panel) and the helical jet model (right panel). In all three cases the line of sight passed parallel to the coordinate Oxy-plane at an aiming distance $x$ from it. \textcolor{black}{In the two-zone jet model $l_1$ and $l_2$ are the projections of the inner radius $\rho_1$ and the outer radius $\rho_2$ on the y-axis.} The model magnetic fields $\textbf B$ are schematically indicated by bold red lines. Detailed descriptions of each model are presented in the sections 2, 3 and 4.}
\label{Fig1} 
\end{center}
\end{figure*}
As we mentioned in the introduction the physical phenomenon underlying Burn's formula is the Faraday rotation, which occurs due to the incoming phase difference between the right and left polarized waves \citep{Ginzburg:1966}. The difference between the refractive indices for these two waves is so small, that the main role here is played by the large distance, over which a rotation of the polarization plane occurs. Therefore the local polarization angle $\psi(r)$ can be determined, see, e.g., \cite{sokoloff1998depolarization}, by the integral expression along the line of sight:
\begin{equation}\label{E1}
\psi(r)=\psi_0(r)+R(r)\lambda^2, \quad\text{where}\quad R(r)=-K \int_{r_0}^{r}n\textbf B d\textbf r.
\end{equation}
\textcolor{black}{The function $\psi_0(r)$ here represents the intrinsic position of the electric vector of the synchrotron emission, which is assumed to be perpendicular to the transverse component of the magnetic field $\textbf B$, $\lambda$ is the  radiation wavelength, $r_0$ and $r$ are the start and end points of the beam passing through the active region, $n$ is the number density of the thermal electrons and the notation ${K}$ is used for the ratio ${e^3}/{2\pi m^2 c^4}$ (in cgs electromagnetic units), where $c$, $e$ and $m$ are the the speed of light, the electron charge and mass correspondingly.}

To calculate the depolarization effect from the extended astrophysical objects, e.g. galaxies and jets, it must be taken into account that each element along the line of sight can have its own emissivity $\varepsilon(r)$ -- the energy emitted towards the observer per unit time per unit volume. Thus, the complex linear polarization $P$ should be defined by the integration along the line of sight:
\begin{equation}\label{E2}
\frac{P}{P_0}={\int_{r_0}^{r}{\varepsilon(r) \exp\left(2i(\psi_0(r)+{R}(r)\lambda^2)\right)}dr}\left|
\int_{r_0}^{r}{\varepsilon(r)dr}\right|^{-1},
\end{equation}
where, following \cite{Ginzburg:1966}, \textcolor{black}{we assume that the emissivity $\varepsilon(r)$ is determined only by the square of the perpendicular component of the magnetic field\footnote{{This assumes that the spectral index of the radiation $\alpha = -1$, where the intensity depends on the frequency $\nu$ as $\nu^{\alpha}$}. \textcolor{black}{Thus, the intrinsic degree of polarization $P_0$ is 0.75}}}. \textcolor{black}{Moreover, here and below we consider only a completely resolved sources, keeping in mind, that finite resolution, e.g., for the Gaussian beams, discussed by \cite{burn1966depolarization} and \cite{sokoloff1998depolarization}, could significantly complicate the analysis \citep[see also][]{1980AJ.....85..368C}.}

Strictly speaking Burn's result itself provided the complex polarization (\ref{E2}) of the synchrotron radiation from a homogeneous plate, \textcolor{black}{which can be used as the approximation of a galaxy disk with a constant thickness $L=2l$ (though Burn did not apply his slab model to a disk galaxy at all)}. Using the coordinate system referenced to the homogeneous plate, where y-axis is directed perpendicular to the 'galaxy' plane and to the 'galaxy' internal magnetic field $\textbf B=(0,0,-B_z)$, we direct the line of sight (the direction opposite to the direction of the emission) at the angle $\theta$ to the z-axis, see the left panel of the Figure \ref{Fig1}. \textcolor{black}{For a constant magnetic field $B_z$ and constant number density $n$ from the equation~(\ref{E1}) it is easy to calculate the function ${R}(y)$, substituting the variable $r$ along the line of sight with the $y$ coordinate along the y-axis and considering $d\textbf r=(0,dy,dy\,{\rm ctg}\theta)$:}
\begin{equation}\label{E3}
{R}(y)=Kn\int_{-l}^{y} B_z{\rm ctg}\theta\, dy= Kn B_z{\rm ctg}\theta \,(l+y).
\end{equation}
{Then, introducing the so called intrinsic Faraday rotation measure $\mathcal{R}=2 Kn B_z{\rm ctg}\theta \,l$, see, e.g., \cite{sokoloff1998depolarization}}, from the equation (\ref{E2}) it is possible to obtain Burn's relation for the complex relative polarization $P/P_0$, more conveniently for its absolute value -- the polarization degree $\Pi$ and its argument -- the polarization angle $\Psi$: 
\begin{multline}\label{E4}
\frac{P}{P_0}= \Pi\exp(2i\Psi)=\frac{\sin(\mathcal{R}\lambda^2)}{\mathcal{R}\lambda^2 }\exp(2i(\psi_0+\mathcal{R} \lambda^2/2)) \, ,\\
%\;\;\Rightarrow \\ \Rightarrow\;\; 
\text{ and } \;\;
\Pi=\frac{\sin(\mathcal{R}\lambda^2)}{\mathcal{R}\lambda^2 } \, ,
\;\;\text{ where }\;\; \Psi=\psi_0+\mathcal{R} \lambda^2/2.
\end{multline}
Thus, according to Burn's formula, the polarization degree $\Pi$ depends on the wavelength squared $\lambda^2$ as a sinc function, and the polarization angle $\Psi$ -- as a linear function with the {slope} coefficient $RM=\mathcal{R}/2$, usually called the Faraday rotation measure. However, it is important to understand that the observed polarization angle $\Psi_{\rm obs}$ follows the linear law except for discontinuities at $\mathcal{R}\lambda^2=n\pi$, where sinc {changes its sign}. At these points the sign of the polarization degree $\Pi$, which is non-negative, does not change, but \textcolor{black}{there is} a jump of the exponent argument by $i\pi$. \textcolor{black}{This jump corresponds to the change of the polarization angle by $\pi/2$ and, as a result, the linear dependence transforms into the well-known sawtooth dependence with amplitude $\pi/2$. Such type of the sawtooth structure also occurs in a more complex field models with the complex polarization having one or more zeros - in particular, for jet magnetic field models discussed below.}

Finally, note that the simple form of Burn's results (\ref{E4}) follows from a number of assumptions that are reasonable only as an initial approximation. One have to keep in mind that the real systems are significantly more complex. However, talking in this paper about AGN jets, we limit ourselves to the \textcolor{black}{similar} assumptions, so it makes sense to list them again. First of all, we assume the constant number density $n$ and the emissivity $\varepsilon(r)$, determined only by the square of the perpendicular component of the magnetic field \textcolor{black}{implying spectral index $\alpha = -1$ and $P_0 = 0.75$}, by analogy with the reasoning of the classical work \cite{Ginzburg:1966}. Second, Burn's formula assumes a constant initial positional angle $\psi_0$, as the angle between the observer's chosen direction ${\bf N}$ on the celestial sphere and the electric wave field, oscillated parallel to the vector product ${\bf B}\times d{\bf r}$. Then the constant angle $\psi_0$ in (\ref{E4}) follows from the constant magnetic field in the galactic disk. We use the same assumption, $\psi_0$ is the angle between the vector ${\bf N}$ and the vector ${\bf B}\times d{\bf r}$, for jets also. Finally, the linear dependence of the polarization angle $\Psi$ on the wavelength squared $\lambda^2$ is a direct consequence of the \textcolor{black}{magnetic field} symmetry with respect to $y=0$, see remarks in Appendix \ref{AppendixA}. For jets we consider the azimuthally symmetric cases leading to similar conclusions. Under more complicated assumptions, this dependence can have much more complex character, see, e.g., \cite{sokoloff1998depolarization}. The discussion of the assumptions correctness in the light of the observed data and the approximation of more complex AGN jet models are left by us for the future works.

%%%%%%%%%%%%%%%%%%%%%%%%%%%%%%%%%%%%%%%%%%%%%%%%%%%%

\section{Two-zone magnetic field model}

\begin{figure}
\begin{center}
\includegraphics[width=9cm]{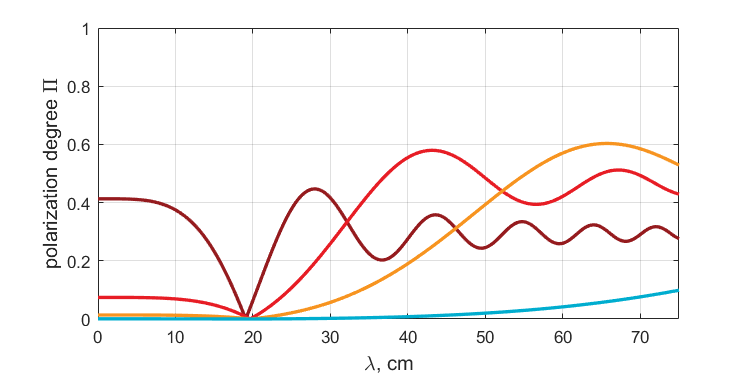}
\includegraphics[width=9cm]{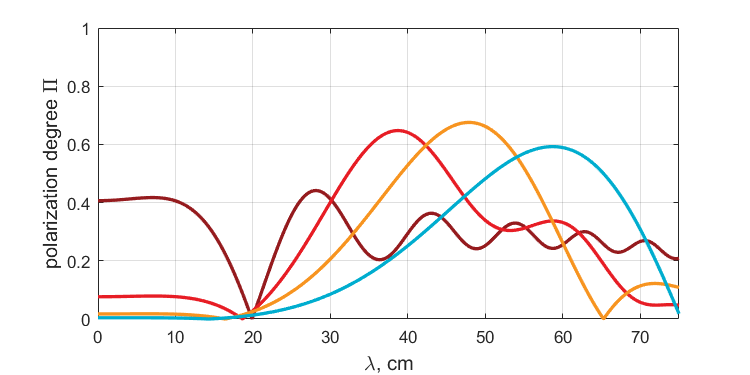}
\includegraphics[width=9cm]{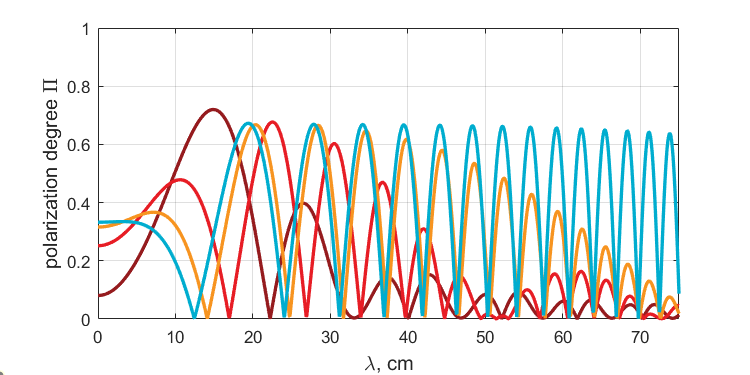}
\caption{Two-zone magnetic field model. Dependencies of the polarization degree $\Pi$ on wavelength $\lambda$ for three angles: $\theta=90^{\circ}$ (top panel), $\theta=85^{\circ}$ (middle panel), and $\theta=45^{\circ}$ (bottom panel). The colors of the lines correspond to different aiming distances: $x=0.05\rho_2$ (blue), $x=0.25\rho_2$ (orange), $x=0.50\rho_2$ (red) and $x=0.75\rho_2$ (brown). The azimuthal and longitudinal magnetic fields here are $KnB_{\varphi}=0.01$ and $KnB_z=0.005$, so the Faraday rotation measure varies from $0$ to about $50$ ${\rm rad/m^2}$ for $\theta=90^{\circ}$, see the equation (\ref{E8}) and Figure \ref{Fig5}.}
\label{Fig3} 
\end{center}
\end{figure}

The derivation of the relation similar to Burn’s formula for cylindrical structures is also possible.
The main question here is how to set up such a structure in such a way that, firstly, the main features of the depolarization can be predicted and, secondly, the simplest and most convenient formula comparable to Burn's relation can be obtained.
As a zeroth \textcolor{black}{order} approximation consider a cylindrical model consisting of two zones: an inner cylinder ($0\le\rho\le\rho_1$) and an outer cylindrical {shell} ($\rho_1\le\rho\le\rho_2$), \textcolor{black}{where as in the traditional cylindrical coordinate system $\rho$ is the Euclidean distance from the z-axis and $\varphi$ is the azimuthal angle between the x-axis and the point projection on the Oxy-plane.} Assume that in the central part the constant longitudinal magnetic field $\textbf B=(0,0,-B_z)$ is directed along the z-axis, and at the periphery the azimuthal field is defined as $\textbf B=(B_{\varphi}\sin\varphi,-B_{\varphi}\cos\varphi,0)$, see the middle panel of Figure~\ref{Fig1}. Note that such structure could be considered as an approximation of the force-free magnetic field, where the pitch-angle of the field increases with distance from the jet axis \citep{2005MNRAS.360..869L,2011MNRAS.415.2081C} or MHD jet model with the ``core'' of the longitudinal magnetic field \citep{2023MNRAS.524.4012B}.

Assume that the line of sight is directed exactly to the centre of the jet at an angle $\theta$ to the z-axis, then the case with $\theta=90^\circ$ can reasonably be called degenerate, since there is no magnetic field component directed along the line of sight. For $\theta\ne90^\circ$ the peripheral region provides only the intensity of the radiation, while the rotation of the polarization plane is performed by the central region, similar to Burn's formula (\ref{E4}). In this case, as shown in Appendix \ref{AppendixB}, the polarization degree and the polarization angle will be equal to
\begin{equation}\label{E5}
\Pi=\frac{1}{1+\gamma}\frac{\sin(\mathcal{R}_1\lambda^2)}{\mathcal{R}_1\lambda^2}-\frac{\gamma}{1+\gamma}\cos\left(\mathcal{R}_1\lambda^2\right)\;\text{ and }\;\Psi=\mathcal{R}_1\lambda^2/2,
\end{equation}
where $\mathcal{R}_1=2 KnB_z {\rm ctg\,}\theta \rho_1$ is the intrinsic Faraday measure and $\gamma=\varepsilon_2(\rho_2-\rho_1) / \varepsilon_1 \rho_1$ characterizes the ratio of intensities at the periphery and in the center. As discussed in the \textcolor{black}{\autoref{sec:burns_relation}}
we assumed that the emissivities are determined only by the square of the perpendicular component of the magnetic field, in other words $\varepsilon_1 =\kappa B_z^2\sin^2\theta$ and $\varepsilon_2 =\kappa B_{\varphi}^2$, where $\kappa$ is the proportionality factor.

It is interesting that even \textcolor{black}{within} such a simple consideration, the difference between the radiation from a flat galaxy (\ref{E4}) and from an AGN jet (\ref{E5}) \textcolor{black}{is evident}. For example, for small $\mathcal{R}_1\lambda^2$, when $\theta$ is {close} to $90^\circ$ or $\lambda$ is {close} to $0$, the synchrotron radiation is initially depolarized: $\Pi(0)= |1-\gamma|/|1+\gamma|\le 1$. The reason for this is that in the central and peripheral regions the electric fields oscillate along the different directions: the electric field is perpendicular to the x-axis at the periphery and parallel to the x-axis at the centre. Thus, the modulus of $\Pi(\lambda^2)$, which for Burn's formula has the form of a sinc, is now the difference between the sinc and the cosine function, where the mutual contribution of the components is determined by $\gamma$. This can be seen in Figure \ref{Fig3}, where the dependencies of the polarization degree $\Pi$ on the wavelength $\lambda$ \textcolor{black}{for different angles, $\theta$, and aiming distances, $x$, are shown. The low value of polarization, called below the polarization drop,} is clearly seen for the jet center lines -- blue lines with $x=0.05\rho_2$ -- for all three cases: $\theta=90^{\circ}$ (top panel), $\theta=85^{\circ}$ (middle panel), and $\theta=45^{\circ}$ (bottom panel). Another example is the dependence of the polarization angle $\Psi$ on the wavelength squared $\lambda^2$. For Burn's relation, this dependence is linear (\ref{E4}) with the polarization angle $\Psi=\psi_0$ for small $\lambda$. Finding this $\psi_0$ is an important \textcolor{black}{observational} problem, which is feasible but requires multiple measurements at different wavelengths. For the jet the angle $\Psi$ has no $\psi_0$ because of the parity of the magnetic field component along the line of sight, see Appendix \ref{AppendixA}. It would be more correct to say that in order for the angle $\Psi=0$ at small $\lambda$, it is necessary to count the \textcolor{black}{polarization position angle} from the direction perpendicular to the axis of the jet in the plane of the sky, x-axis. Note also that the actual observed polarization angle $\Psi_{\rm obs}$ dependence is not a linear, but sawtooth with discontinuities of $90^{\circ}$, where formula (\ref{E5}) for $\Pi$ changes sign\footnote{The example which shows the typical sawtooth dependence of $\Psi(\lambda^2)$ \textcolor{black}{for helical magnetic field \autoref{sec:from_center_to_edge}} is shown in Figure \ref{Fig2}. \textcolor{black}{For two-zone model the dependence is similar.}}. \textcolor{black}{Especially interesting is the situation at small wavelengths, when the $\pi/2$ jump can also be at $\lambda=0$. For example, in the case when formula (\ref{E5}) gives a negative value for $(1-\gamma)/(1+\gamma)$. In other words, at small wavelengths the observed polarization angle $\Psi_{\rm obs}=0$, if $\gamma<1$, and $\Psi_{\rm obs}=\pi/2$, if $\gamma>1$.}

Now consider the line of sight that does not pass through the centre, but is parallel to the Oyz-plane  at a distance of $x$ from the axis of the jet. Suppose that this aiming distance $x$ is small enough that the line still passes through two zones: the central one with a field $B_z$ parallel to the jet axis and the peripheral one with a constant azimuthal field $B_{\varphi}$. Let's neglect the change of the azimuthal field rotation along line of sight in the outer cylindrical ring, i.e. assume that the angle between the magnetic field and the line at periphery does not change and is equal to $\varphi_0$: $\cos\varphi_0=x/\rho_2$. Calculations similar to those made in the \textcolor{black}{\autoref{sec:burns_relation}}, see Appendix \ref{AppendixB}, allow us to calculate the complex polarization $P/P_0$. Since the magnetic field $\bm{B}(y)$ along the line of sight turns out to be an even function with respect to $y=0$, and the initial position angle $\psi_0(y)$ on the contrary is an odd function (due to the rotation of the azimuthal magnetic field), the polarization angle $\Psi$ again turns out to be a linear function of the wavelength squared, see Appendix \ref{AppendixA}:
\begin{equation}
\Psi=\left(\mathcal{R}_2+\mathcal{R}_1/2\right)\lambda^2.
\end{equation}
The slope of this linear dependence is determined by the difference of two intrinsic Faraday measures corresponding to the central and peripheral parts:
\begin{equation}\label{E7}
\mathcal{R}_1=2 Kn B_z {\rm ctg\,}\theta\, l_1 \;\;\text{ and }\;\;\mathcal{R}_2=Kn B_{\varphi}\,x(l_2-l_1)/\rho_2,    
\end{equation}
where $l_1=(\rho_1^2-x^2)^{1/2}$ and $l_2=(\rho_2^2-x^2)^{1/2}$.
It is clearly seen that the Faraday rotation measure $RM=\mathcal{R}_2+\mathcal{R}_1/2$ changes with both the angle $\theta$ and the aiming distance $x$ (see, e.g., the middle panel of Figure \ref{Fig5}). This implies that it is possible to reconstruct the internal structure of the jet magnetic field by measuring this slope for different $x$. Such an inverse problem is interesting in itself, but we leave it for the next paper and limit ourselves to noting that it can be done in principle. 

Similarly, as shown in Appendix \ref{AppendixB}, the expression for the degree of polarization $\Pi$ can be obtained:
\begin{multline}\label{E8}
\Pi=\frac{1}{1+\gamma}\frac{\sin(\mathcal{R}_1\lambda^2)}{\mathcal{R}_1\lambda^2}+\frac{\gamma}{1+\gamma}\frac{\sin(\mathcal{R}_2\lambda^2)}{\mathcal{R}_2\lambda^2}\cos\left((\mathcal{R}_2+\mathcal{R}_1)\lambda^2+2\psi_0\right),\\
\text{where }\quad \gamma=\frac{\varepsilon_2 (l_2-l_1)}{\varepsilon_1 l_1}=\frac{B_{\varphi}^2}{B_z^2}\frac{\rho_2^2-x^2\sin^2\theta}{\rho_2^2\sin^2\theta}\frac{l_2-l_1}{l_1}
\end{multline}
It is seen that, as well as for the line of sight incident to the center, the \textcolor{black}{polarization degree} is a superposition of the sine and cosine functions\textcolor{black}{. Their ratio} is determined by the the relative emissivity $\gamma$ of the central and peripheral regions. Besides, \textcolor{black}{there is} an additional term in cosine's argument $\psi_0$: $\cos\psi_0=x\cos{\theta}/\sqrt{\rho_2^2-x^2\sin^2{\theta}}$, close to $\pi/2$ at small aiming distances $x$, which gives a difference of sine and cosine, and hence a reduced the polarization degree $\Pi(0)=|1-\gamma|/|1+\gamma|$ at small values of $\lambda$.

Figure~\ref{Fig3} \textcolor{black}{shows the results for the equation (\ref{E8})} for three angles $\theta$ at different panels, \textcolor{black}{revealing} a clear difference from Burn's relation (\ref{E4}) -- the maximum value of the polarization degree does not correspond to small wavelengths. The degenerate case shown in the top panel has another difference mentioned earlier -- the polarisation may not tend to zero at long wavelengths, but to some constant level. {However, already at \textcolor{black}{angles close to $\pi/2$}, e.g. the middle panel of Figure~\ref{Fig3},} this situation is different and the polarization degree {converges}, albeit slowly, to zero with increasing $\lambda$. A similar situation is observed at small aiming distances{, that corresponds to the blue lines in Figure~\ref{Fig3}}. As mentioned in Appendix \ref{AppendixB}, it is possible that the polarization degree oscillates like a trigonometric function without decreasing. In the general case it can be stated that the \textcolor{black}{observed} polarisation of jet radiation will be more difficult to analyse. First, because of the reduced degree: for a large class of parameters there are no values of the polarization degree (\ref{E2}) close to unity. And second, because of the difficulty in predicting at which wavelengths the maximally polarised radiation will be observed.

%%%%%%%%%%%%%%%%%%%%%%%%%%%%%%%%%%%%%%%%%%%%%%%%%%%%%%%%%%%%%%%%%%%%%%%%%%%
\section{Helical magnetic field model}
\label{sec:from_center_to_edge}
\begin{figure}
\begin{center}
\includegraphics[width=9cm]{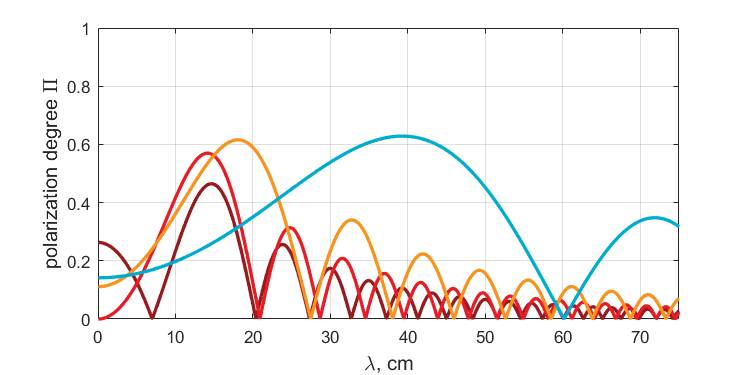}
\includegraphics[width=9cm]{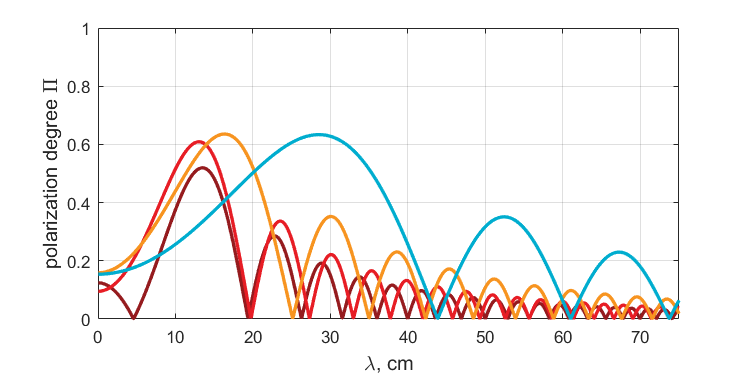}
\includegraphics[width=9cm]{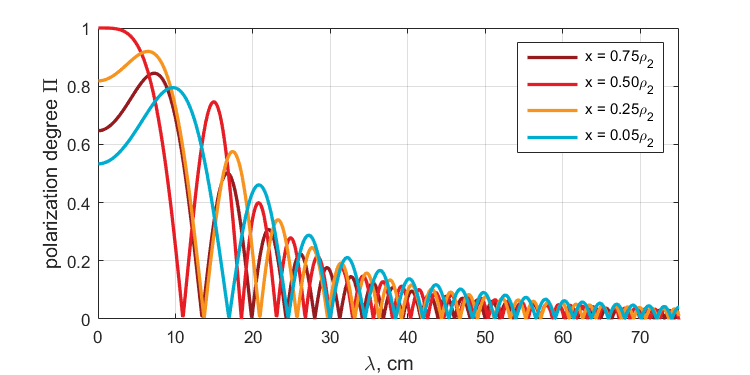}
\caption{Helical {magnetic field} model. Dependencies of the polarization degree $\Pi$ on wavelength $\lambda$ for three \textcolor{black}{viewing} angles: $\theta=90^{\circ}$ (top panel), $\theta=85^{\circ}$ (middle panel), and $\theta=45^{\circ}$ (bottom panel). The colors of the lines correspond to different aiming distances: $x=0.05\rho_2$ (blue), $x=0.25\rho_2$ (orange), $x=0.50\rho_2$ (red) and $x=0.75\rho_2$ (brown). The azimuthal and longitudinal magnetic fields are $KnB_{\varphi}=0.01$ and $KnB_z=0.005$ that the internal Faraday measures vary from $0$ \textcolor{black}{up to} about $50$ ${\rm rad/m^2}$ for $\theta=90^{\circ}$, see the equation (\ref{E12}) and Figure \ref{Fig5}.}
\label{Fig4} 
\end{center}
\end{figure}

\begin{figure}
\begin{center}
\includegraphics[width=9cm]{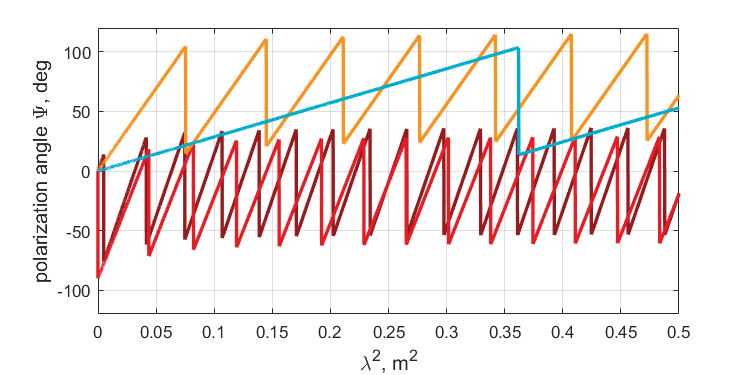}
\caption{Typical dependence of the polarization angle $\Psi$ on the square of the wavelength $\lambda^2$ for the \textcolor{black}{helical} magnetic fields model, $\theta=90^{\circ}$. The polarization angle is counted from the x-axis shown in Figure \ref{Fig1} and corresponding to the direction perpendicular to the projection of the jet axis onto the plane of the sky, see Appendix \ref{AppendixA}. The parameters and colors here are the same as in the top panel of Figure \ref{Fig4}. The sawtooth form is associated with the sign change of the formula for the polarization degree $\Pi$.}
\label{Fig2} 
\end{center}
\end{figure}

The two-zone model has highlighted the main differences between the polarization of radiation from galaxies and jets. Apart from the polarization drop at a short wavelengths and the direct proportionality between the polarization angle and the wavelength squared, the dependence of the polarization properties on the aiming distance $x$ seems to be the most interesting. This is due to the axial symmetry of the jet and the fact that the aiming distance uniquely determines the region through which the radiation has passed. However, the two-zone model is inappropriate for a correct study of the aiming distance $x$ dependence of the depolarization, since this model contains the separation boundary between the longitudinal field $B_z$ and the azimuthal field $B_{\varphi}$. So it worth to consider another jet model - with a helical magnetic field. 

\begin{figure}
\begin{center}
\includegraphics[width=9cm]{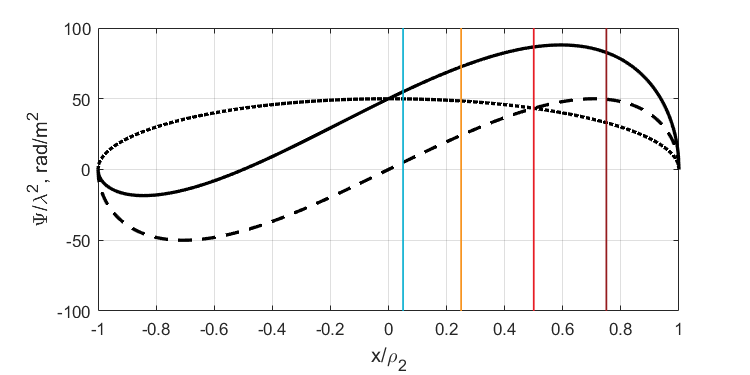}
\includegraphics[width=9cm]{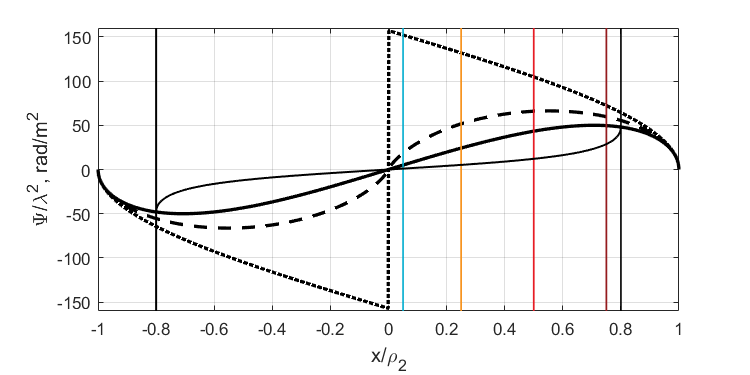}
\includegraphics[width=9cm]{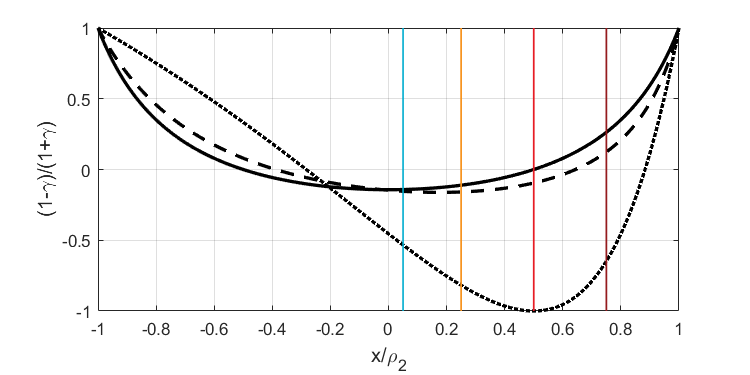}
\caption{Top panel shows the dependencies of the Faraday rotation measures $\Psi/\lambda^2$ on the aiming distance $x/\rho_2$ for the helical field at angle $\theta=90^\circ$ (dashed line) and at angle $\theta=45^\circ$ (solid line). At angle $\theta=90^\circ$ the curve is odd relative to the jet axis. At angle $\theta=45^\circ$ the curve is a superposition of the even component (dotted line) and the odd component (coincides with the dashed line). The middle panel shows the dependencies of the Faraday rotation measures $\Psi/\lambda^2$ on the aiming distance $x/\rho_2$ for different magnetic fields: for the increasing field (solid line), for the constant field (dashed line), for the decreasing field (dotted line) and also for the two-zone model (thin solid line, for $-0.8<x/\rho_2<0.8$). The bottom panel shows the dependencies of the ratio $(1-\gamma)/(1+\gamma)$ on the aiming distance $x/\rho_2$, the modulus of which determines the degree of polarization at small wavelengths for different angles: $\theta=90^\circ$ (solid line), $\theta=85^\circ$ (dashed line) and $\theta=45^\circ$ (dotted line). The azimuthal and longitudinal magnetic fields are the same as in Figures \ref{Fig3} and \ref{Fig4}, $KnB_{\varphi}=0.01$ and $KnB_z=0.005$}
\label{Fig5} 
\end{center}
\end{figure}

Consider a cylinder of radius $\rho_2$ extended along the z-axis, see the right panel of Figure \ref{Fig1}, with the helical magnetic field $\textbf B$:
\begin{equation}
\bm{B} = (B_{\varphi}\sin{\varphi}(y), -B_{\varphi}\cos{\varphi}(y), -B_z),
\end{equation}
where the angle $\varphi(y)$ now depends not only {on} the aiming distance $x$, but also on $y$-coordinate, as $\cos{\varphi} = x/(x^2+y^2)^{1/2}$, where $y\in[-l_2,l_2]$ and $l_2=(\rho_2^2-x^2)^{1/2}$. Again, due to the axial symmetry, such a magnetic field along the line of sight is defined by the even function, so all the results of Appendix \ref{AppendixA} are still applicable. Appendix \ref{AppendixC} presents the example of the complex polarization calculation for a line of sight passing at the aiming distance $x$ from the jet axis. Of course, the specific relations for the polarization angle and the degree of polarization depend on the radial profile of the magnetic field components\textcolor{black}{. However} due to the symmetry of the field, the polarization angle turns out to be proportional to the wavelength squared in all cases considered: $\Psi=RM\lambda^2$, see Figure~\ref{Fig2}. In Appendix \ref{AppendixC} we describe three cases: (A) with a {toroidal} magnetic field component growing with distance to the jet axis $B_{\varphi}(\rho)=B_{\varphi}\rho/\rho_2$ and $B_z={\rm const}$; (B) with a constant magnetic field $B_{\varphi}={\rm const}$ and $B_z={\rm const}$; and (C) with the {toroidal} component decreasing with distance from the center $B_{\varphi}(\rho)=B_{\varphi}\rho_2/\rho$ and $B_z={\rm const}$. For all three cases, the Faraday rotation measure can be computed explicitly:
\begin{equation} 
\begin{array}{lll} 
\text{(A):}\quad RM=Kn B_{\varphi}\,x {l_2}/{\rho_2}+Kn B_z{\rm ctg}\theta l_2,\\
\text{(B):}\quad RM=Kn B_{\varphi} x\ln|(\rho_2+l_2)/x|+Kn B_z{\rm ctg}\theta l_2,\\
\text{(C):}\quad RM=Kn B_{\varphi} \rho_2 {\rm arctg}\left({l_2}/{x}\right)+Kn B_z{\rm ctg}\theta l_2,
\end{array}
\end{equation}
but it looks the simplest for the first case (A), which will be discussed further. The dependencies of $RM$ on the aiming distance $x/\rho_2$ are shown in Figure \ref{Fig5}. Note that the Faraday rotation measure can be represented as the sum of two intrinsic measures:
\begin{equation}\label{E11}
\frac{\Psi}{\lambda^2}=RM=\frac{\mathcal{R}}{2}=\mathcal{R}_2+\frac{\mathcal{R}_1}{2}= Kn B_{\varphi} \frac{x l_2}{\rho_2}+Kn B_z{\rm ctg}\,\theta l_2,    
\end{equation}
\textcolor{black}{which are determined by the longitudinal $B_z$ and azimuthal $B_{\varphi}$ field components.}
% where the first intrinsic measure is determined by the longitudinal field $B_z$ and the second intrinsic measure is determined by the azimuthal field $B_{\varphi}$.
The first is an even $x$-function with respect to the jet center and the second is an odd one, resulting in an asymmetric profile of the Faraday rotation measure with respect to $x=0$. The profile of such an asymmetric dependence, case (A), for the angle $\theta=45^\circ$ is shown in the top panel of Figure \ref{Fig5} (solid line), as well as its symmetric (dotted line) and antisymmetric (dashed line) components. As the angle $\theta$ approaches $90^\circ$, the even component disappears, leaving only the odd component (dashed line). Thus, the symmetric and asymmetric parts of the \textcolor{black}{observationally} obtained profiles allow us to estimate the {relative strength} of the longitudinal and azimuthal magnetic fields. The middle panel shows the Faraday rotation measure profiles for $\theta=90^{\circ}$ and for different magnetic field distributions: the increasing field (case A, solid line), the constant field (case B, dashed line), and the decreasing field (case C, dotted line). Since the longitudinal magnetic field in these cases does not rotate the polarization plane, the profiles are odd with respect to $x = 0$, but at the same time \textcolor{black}{they} have a different structure. This implies that the distribution of the azimuthal magnetic field in jet-like structures can be reconstructed from the similar {observed} profiles. Note again that the discussed rotation measure $RM$ determines the slope of the dependence $\Psi (\lambda^2)$. For example, the middle panel of Figure \ref{Fig5} shows that $RM$, and hence the slope of the dependence, increases toward the edge of the jet, from the blue line ($x=0. 05$) to the brown line ($x=0.75$). This is also seen in Figure \ref{Fig2}, where the slope increases from the blue to the brown line -- but the dependence itself is not linear, but saw-toothed. As mentioned earlier, the jumps in the {linear} dependence occur at the {wavelengths} where the sign of $\Pi$ changes. Thus, the sawtooth line oscillates in the band of $90^{\circ}$, but at different levels. These different levels are determined by the first sign change of $\Pi$. For example, Figure~\ref{Fig2} shows that the \textcolor{black}{for} the red line ($x=0.5$) \textcolor{black}{the sign changes} almost immediately, \textcolor{black}{while for} the blue line ($x=0.05$) sign changes later than \textcolor{black}{for} the others.

The degree of polarization $\Pi$, see Appendix \ref{AppendixC}, also depends on the magnetic field profiles. \textcolor{black}{For the most calculationally convenient case (A), when the field grows from the center of the jet,} the degree of polarization can be written as 
\begin{equation}\label{E12}
\Pi=\frac{1}{1+\gamma}\left(1+\sqrt{3\gamma}\frac{\partial}{\partial \xi}\right)^2\frac{\sin\xi}{\xi},\;\text{ where }\;\xi=\mathcal{R}\lambda^2,
\end{equation}
and $\gamma$ \textcolor{black}{is a more complicated structure than simply the ratio of the squared field components:}
\begin{equation}\label{E13}
\gamma=\frac{l_2^2 B_{\varphi}^2}{3\rho_2^2(B_z\sin\theta-x B_{\varphi}\cos\theta /\rho_2)^2}.
\end{equation}

Figure~\ref{Fig4} demonstrates the dependencies of the polarization degree $\Pi$ on the wavelength $\lambda$ for three angles: $\theta=90^{\circ}$ (top panel), $\theta=85^{\circ}$ (middle panel), and $\theta=45^{\circ}$ (bottom panel). The colors of the lines correspond to different aiming distances: $x=0.05\rho_2$, $x=0.25\rho_2$, $x=0.50\rho_2$, and $x=0.75\rho_2$, which are chosen to be the same as for the two-zone field model. The dependencies generally look similar to those for the two-region model shown in Figure~\ref{Fig3}. However, there are no cases of non-decreasing polarization degree at long wavelengths. This can be seen from the small difference between the upper and middle panels for the rays near the center of the jet $x \approx 0$, the blue lines. There are also no oscillations with a constant amplitude, i.e. the degree of polarization decreases rapidly with increasing wavelength.

Finally, let's note another similarity between Figures~\ref{Fig4} and \ref{Fig3} -- the depression of the polarization degree at small wavelengths. \textcolor{black}{Incidentally, to demonstrate the behavior of the polarization degree at small wavelengths, we plot the dependencies $\Pi(\lambda)$, rather than the more familiar dependencies $\Pi(\lambda^2)$}. In Appendix \ref{AppendixC} \textcolor{black}{we obtain} the following expression for the degree of polarization at short wavelengths $\Pi(0)=|1-\gamma|/|1+\gamma|,$ where
$$\gamma={\int_{0}^{l_2} B_{\varphi}^2\sin^2\varphi(y)\,dy}\left|{\int_{0}^{l_2}(B_z\sin\theta-B_{\varphi}\cos\varphi(y)\cos\theta)^2\,dy}\right|^{-1}$$
Thus such polarization drop has an even profile with respect to \textcolor{black}{the jet axis} for $\theta=90^\circ$, and odd for other angles, see the bottom panel of Figure~\ref{Fig5}.
% Since $\gamma$ determines the relative strength of the longitudinal and azimuthal field components, analysing the {observed} profiles of such a $\Pi(0)$ one can also, as well as on the Faraday rotation measure, set the problem of reconstruction of the jet magnetic field structure.
\textcolor{black}{Furthermore, in this panel we can clearly see that there are regions in the jet with $(1-\gamma)/(1+\gamma)<0$ (near the centre) and $(1-\gamma)/(1+\gamma)>0$ (near the periphery). As mentioned in \autoref{sec:burns_relation}, this means that at small wavelengths the observed polarization is either transverse $\Psi_{\rm obs}=0$ ($\gamma<1$) or longitudinal $\Psi_{\rm obs}=\pi/2$ ($\gamma>1$) due to the jet symmetric structure, see Appendix \ref{AppendixA}.}
Since $\gamma$ determines the relative strength of the longitudinal and azimuthal field components, the {observed} profiles of $\Pi(0)$ can also, as well as the Faraday rotation measure profiles, constrain the jet magnetic field structure. However, this problem is beyond the scope of this study and will be presented in the next paper.

\section{Discussion}
% Relation to previous works.
It should be noted, that the ``inverse (or anomalous) depolarization'' was \textcolor{black}{considered} earlier in the context of the multifrequency polarimetry of spiral galaxies \citep{sokoloff1998depolarization} and AGN jets with helical magnetic fields \citep{2012ApJ...747L..24H}. \textcolor{black}{\cite{2021MNRAS.507.4968F} studied the general case of the magnetic field with a helicity and confirmed the influence of the helical magnetic field on the Faraday rotation.} In \cite{2012ApJ...747L..24H} the problem was investigated numerically by solving the full equations of the radiative transfer for a cylindrical jet with the magnetic field similar to the one described in Section~\ref{sec:from_center_to_edge}. In our paper we present the analytical treatment for a wider range of magnetic field models. Also \citet{2013MNRAS.430.1504M} applied analytical expressions of the transverse polarization profiles for the helical magnetic field from \citet{1981ApJ...248...87L} to assessing the magnetic field pitch-angle and the viewing angle (both in the emitting plasma frame) for the parsec jet of Mrk~501, successfully recovering the spine-sheath structure. However, they haven't considered the internal Faraday rotation and focused solely on the helical magnetic field influence on the polarization profiles at a single frequency.

% How we bypass the relativistic motion consideration.
In our models we do not consider the relativistic bulk motion of the jet. AGN jets at parsec scales are thought to have a highly relativistic speeds \citep{2016AJ....152...12L,2019ApJ...874...43L,2021ApJ...923...30L,2022ApJS..260...12W}{, while for kpc-scale jets the bulk motion is sub or mildly relativistic \citep{1997MNRAS.286..425W,1997ApJ...488..675W,2004MNRAS.351..727A,2009MNRAS.398.1989M,2016ApJ...818..195M,2017Galax...5....8M}}. However, as discussed in
\citet{2013MNRAS.430.1504M}, the relativistic bulk motion induces Doppler boosting and aberration, which do not change the polarization profiles, provided that the jet's velocity is constant across its width. Thus, similar to \citet{2013MNRAS.430.1504M} we effectively consider the emission and rotation in the rest frame of the moving plasma. Jets from the flux limited samples have viewing angles $\approx 1/\Gamma$ \citep{1994ApJ...430..467V,2019ApJ...874...43L}. It implies that in the plasma frame they are viewed nearly ``side-on'' \citep{2021Galax...9...58G}. Another consequence of the relativistic bulk motion, as noted by \citet{2012ApJ...747L..24H}, is that in the comoving frame the plasma sees radiation with larger, $\propto D/(1 + z)$, where $D$ - Doppler factor of the Faraday rotating plasma bulk motion, $z$ - redshift of the source, than the observed wavelength due to the Doppler effect. This implies that the internal rotation is enhanced even if the rotating particles are \textcolor{black}{not thermal, but} mildly relativistic, \textcolor{black}{e.g. from the low end of the power-law energy distribution.}

We stress, that in our models the transverse gradients of $RM$ trace the jet magnetic field itself, not the magnetized sheath around the jet \citep[e.g.][]{2010ApJ...725..750B,2011MNRAS.415.2081C,2013MNRAS.436.3341Z}. Actually, this does not contradict to the previous studies of the blazar jets. The typical $RM$s \textcolor{black}{observed in parsec-scale} jets are quite modest \citep[$\approx100$ rad/m$^2$ in 8--15 GHz band with 80$\%$ of the jets in the MOJAVE sample showing $RM \leq$ 400 rad/m$^2$,][]{1998ApJ...506..637T,2000ApJ...533...95T,2001ApJ...550L.147Z,2012AJ....144..105H}. Even \textcolor{black}{the} larger values of $RM$ ($\approx$ 1000 rad/m$^2$) observed in some jets \citep[e.g. between 12 and 43 GHz in][]{2013MNRAS.429.3551A} \textcolor{black}{are not accompanied with large enough rotation to exclude the internal case.} Moreover, depolarization of some sources as well as the observed circular polarization do require the internal rotation \citep{2009ApJ...696..328H,2012AJ....144..105H,2017MNRAS.467...83K}. \textcolor{black}{It is} interesting that \citet{2012ApJ...747L..24H} considers the inverse depolarization and the internal rotation in a jet with a helical magnetic field. {Strictly speaking, inverse depolarization is not a unique signature of the internal rotation: it can also occur for a partially resolved foreground Faraday screens if two emitting regions within the beam have different intrinsic polarization directions and different values of $RM$. However, as noted in \cite{2012ApJ...747L..24H}, jet features displaying this effect are reasonably isolated from other jet features and almost all of them reside in a jets with a transverse $RM$ gradient \citep{2012AJ....144..105H}.} This could imply that the helical magnetic field drives both effects.
We note that some observations clearly indicate the contribution of the external Faraday rotation because of the large polarization angle rotation proportional to the wavelength squared or the variability of $RM$ \citep{2000ApJ...533...95T,2001ApJ...550L.147Z,2009ApJ...694.1485K,2011ApJ...733...11G,2017MNRAS.467...83K}, although some smooth variability could not exclude the rotation by the jet plasma \citep{2009MNRAS.393..429O}. \textcolor{black}{Another simple way to differentiate the transverse gradient of $RM$ driven by the jet magnetic field from that generated in the jet sheath is to consider $RM$ values at the edges of the jet. In models considered in this paper they should be zero because of the zero path length \citep{1981ApJ...248...87L}. However, the finite resolution, noise and contribution from the external screen could distort this simple picture.}

\section{Conclusions}

The main conclusions of the paper, which is our first look at the synchrotron emission depolarization in AGN jets, can be summarized in two points:

\noindent 1. The axial symmetry of the magnetic field leads to a linear dependence of the \textcolor{black}{polarization position} angle on the wavelength squared. The proportionality coefficient depends on the aiming distance and allows to estimate the radial profiles of \textcolor{black}{helical magnetic field components.} Solving this inverse problem in general case will be done in the forthcoming paper. 

\noindent 2. The presence of both the toroidal and longitudinal components of the magnetic field could result in the degree of the polarization \textcolor{black}{having its maximum value at $\lambda > 0$} - contrary to Burn's relation. In other words, the dependence of the polarization degree on the wavelength squared is not just a sinc-function. \textcolor{black}{We derived the relations for the polarization degree transverse profiles for several magnetic field models, generalizing earlier results to the case of the internal Faraday depolarization.}

\textcolor{black}{Finally, we note that all formulas obtained in this paper make the reasonable assumption that jets are axisymmetric.} This implies the proportionality of the \textcolor{black}{polarization position} angle and the wavelength squared. It is also interesting that the dependency of the polarization on the aiming distance allows \textcolor{black}{us} to consider the inverse problem of reconstructing the magnetic field profiles across the jet. \textcolor{black}{From the mathematical point of view, the inference of the magnetic field profile requires a monotonic relation between the coordinate along the ray and the Faraday depth as well as axisymmetry.} Therefore, although the proposed method \textcolor{black}{cannot} be used to reconstruct an arbitrary magnetic field structure, with the assumption of the unidirectional azimuthal field component\footnote{However, some models violate this assumption \citep[e.g.][]{2013MNRAS.431..695M}} the solution of the inverse problem seems quite feasible. Moreover, in the next paper this inverse problem will be solved completely through the solution of the Volterra integral equation.

\section*{Acknowledgements}
\textcolor{black}{We thank the anonymous referee for helpful comments and suggestions that significantly improved the presentation of the results. We thank Naga Varun for careful reading the manuscript and valuable advices.} We are grateful to Dr. R. Beck (MPIfR, Bonn) for stimulating discussions and ideas.

\section*{Data Availability}

There is no new data associated with the results presented in
the paper. All the previously published data has the proper
references.



\bibliographystyle{mnras}
\bibliography{2024Sokoloff} % if your bibtex file is called example.bib

%%%%%%%%%%%%%%%%% APPENDICES %%%%%%%%%%%%%%%%%%%%%

\appendix

\section{Proportionality of polarization position angle to the wavelength squared}\label{AppendixA}
Consider the polarization {for the} magnetic field structures symmetric about the Oxz-plane ($y=0$), exactly such structures are used in this work and shown in three panels of Figure~\ref{Fig1}. Rewrite the numerator $I$ of the complex polarization ratio (\ref{E2}) using as integration variable the coordinate $y=r\sin\theta$ varying on the segment $[-l,l]$:
\begin{equation}\label{A1}
I=\int_{-l}^{l}{\varepsilon(y) \exp\left(2i\left(\psi_0(y)+ {R}(y)\lambda^2\right)\right)}\left(\frac{dy}{\sin\theta}\right),
\end{equation}
where all the functions included in the integral are expressed through the variable $y$. Assume that the emissivity function $\varepsilon(y)$ is even and reduce $I$ to the integration over the segment $[0,l]$ by dividing the integral in the right side (\ref{A1}) in two parts and substituting the variables $y\to -y$ in one of them. Thus we obtain 
\begin{multline}\label{A2}
I=\int_{0}^{l} \varepsilon(y) \left(\exp(2i(\psi_0(y)+R(y)\lambda^2))+\right.\\\left.+\exp(2i(\psi_0(-y)+R(-y)\lambda^2))\right)\left(\frac{dy}{\sin\theta}\right).   
\end{multline}
Rewrite the functions $R(y)$ and $R(-y)$ as
\begin{equation}
R(y)=Kn \int_{-l}^{0} B_{\parallel}({y})\left(\frac{dy}{\sin\theta}\right)+Kn \int_{0}^{y}B_{\parallel}({y})\left(\frac{dy}{\sin\theta}\right),
\end{equation}
\begin{multline}
R(-y)=Kn \int_{-l}^{-y} B_{\parallel}({y})\left(\frac{dy}{\sin\theta}\right)=Kn \int_{y}^{l}B_{\parallel}(-{y})\left(\frac{dy}{\sin\theta}\right)=\\=Kn \int_{0}^{l}B_{\parallel}(-{y})\left(\frac{dy}{\sin\theta}\right)-Kn \int_{0}^{y}B_{\parallel}(-{y})\left(\frac{dy}{\sin\theta}\right),
\end{multline}
where $B_{\parallel}(y)dy$ is the magnetic field component parallel to the line of sight $-\textbf B d\textbf r/|d\textbf r|$ expressed through $y$. For convenience introduce a substitution called the Faraday depth in \cite{burn1966depolarization}:
\begin{equation}
\phi(y)=Kn \int_{0}^{y}B_{\parallel}({y})\left(\frac{dy}{\sin\theta}\right).
\end{equation}
{Assume that the function $B_{\parallel}(y)$ is even and the function $\psi_0(y)$ is odd. This assumption applies, for example, to the cylindrical regions considered in Figure~\ref{Fig1}, if the angle $\psi_0(y)$ is zero on the x-axis.} Rewrite the numerator $I$ in (\ref{A2}) as
\begin{equation*}
I=\int_{0}^{l} 2\varepsilon(y) \cos\left(2\psi_0(y)+2\phi(y)\lambda^2\right)\left(\frac{dy}{\sin\theta}\right) \exp\left(2i\phi(l)\lambda^2\right).
\end{equation*}
Substituting $I$ in the ratio (\ref{E2}), we obtain formulas for the polarization angle $\Psi$ and polarization degree $\Pi$. Moreover, for the angle we immediately obtain the proportional dependence $\Psi=\phi(l)\lambda^2$ with Faraday rotation measure equal $RM=\phi(l)$, and for the polarization degree, having done the same with the denominator of (\ref{E2}), we get
\begin{equation}\label{A6}
\Pi=\int_{0}^{l} \varepsilon(y) \cos\left(2\psi_0(y)+2\phi(y)\lambda^2\right)dy\left|\,
\int_{0}^{l}{\varepsilon(y)dy}\,\right|^{-1}.
\end{equation}
Note once again that the direct proportionality $\Psi=\phi(l)\lambda^2$ is just a direct consequence of the parity of functions $\varepsilon(y)$ and $B_{\parallel}(y)$ and the oddity of function $\psi_0(y)$, i.e. their symmetry with respect to the Oxz-plane. {For Burn's formula, there is also symmetry, see the left panel of Figure~\ref{Fig1}. However, there is no oddness of the function $\psi_0(y)$: in the case of a constant field in the layer this function is equal to a constant. Thus, the dependence of the polarisation angle on the wavelength squared becomes linear $\Psi=\psi_0+\phi(l)\lambda^2$, while the polarization degree, on the contrary, does not include $\psi_0(y)$ and, after taking the integral, results in the famous sinc law (\ref{E4}).}

\section{Two-zone magnetic field model}\label{AppendixB}
Consider the model of a jet consisting of two parts: the central internal part, $0\le\rho\le\rho_1$, with constant magnetic field $\textbf B=(0, 0, -B_z)$ directed along the cylinder axis, and the periphery part, $\rho_1\le\rho\le\rho_2$,  with constant azimuthal magnetic field $\textbf B=(B_{\varphi}\sin{\varphi}, -B_{\varphi}\cos{\varphi}, 0)$. The schematic representation of the areas are shown at the middle panel of Figure~\ref{Fig1}. Suppose that the line of sight passes parallel to the Oyz-plane at the aiming distance $x$ from the jet axis. If we introduce notations $l_1=(\rho_1^2-x^2)^{1/2}$ and $ l_2=(\rho_2^2-x^2)^{1/2}$, {then} the peripheral region corresponds to the range $y\in[-l_2,-l_1] \bigcup [l_1,l_2]$, and the central region to the range $y\in[-l_1,l_1]$. \textcolor{black}{Let's also assume that the change in the angle between the magnetic field $\textbf B$ and the line of sight $d\textbf r$ from $y$ coordinate can be neglected, so this angle $\varphi$ is not defined by the equation $\cos{\varphi} = x/\rho$, but it is always equal to $\pm\varphi_0$, where $\cos{\varphi_0} = x/\rho_2$. This implies that either the aiming distance is small or the periphery sheath is thin. In the first case, if the aiming distance $x$ is small -- for example, the ratio $x/\rho_2$ does not exceed $20\%$ -- the angle $\varphi$ is close to $\pi/2$ both here and there and the error in the angle does not exceed $15\%$. In the second case, if the ratio $(\rho_2-\rho_1)/\rho_2$ is small -- for example, this ratio does not exceed $20\%$ -- then the two cases are also close and the error between $x/\rho_2$ and $x/\rho$ does not exceed $20\%$, and accordingly the angles $\varphi$ and $\varphi_0$ are close again. If both cases are true: $x$ is small and $(\rho_2-\rho_1)/\rho_2$ is small, then the accuracy of the used assumption is even greater and the error is even smaller.}

The intrinsic polarization angle $\psi_0(r)$ is calculated as the angle between the northward direction $\bm{N} = (1,0,0)$ and the electric vector, estimated as the vector product of the magnetic filed $\bm{B}$ and the radiation direction vector $d\bm{r}=(0,dy,dy\,{\rm ctg}\theta)$, thus the its cosine can by calculated as $\cos \psi_0 = \bm{N}(\bm{B}\times d\bm{r})/|\bm{B}\times d\bm{r}|$. Therefore we obtain $\psi_{0}=0$ for the central area $y\in(-l_1,l_1)$ and
\begin{equation}
\cos\psi_0=-x\cos{\theta}/\sqrt{\rho_2^2-x^2\sin^2{\theta}}   
\end{equation}
for the peripheral area. {In the degenerate case of $\theta=90^\circ$ we get $\cos\psi_0=0$, but it is clear that since in the peripheral region the magnetic field vector $\textbf B$ has changed to the opposite one, than the vector $\textbf B\times d\textbf r$ has changed to the opposite one, so we can consider that $\psi_0(y)=\pi/2$ for $y<0$ and $\psi_0(y)=-\pi/2$ for $y>0$. In other words, we can consider the intrinsic polarization angle $\psi_0$ as odd function with respect to $y=0$. Note that the angle $\psi_0(y)$ itself is \textcolor{black}{defined to within an additive constant $n\pi$}, since it is in the argument of the exponent after the multiplier $2i$, see the formula (\ref{E2}). So at any point we can add $\pi$ or remove it, i.e. change the sign of $\cos\psi_0(y)$ - mathematically it does not matter. This gives us a significant advantage when constructing an odd function from $\psi_0(y)$, see the Appendix \ref{AppendixA}. We will not use it in our cases -- because we get vector $\textbf B\times d\textbf r$ with odd $y$ and $z$ components -- but in other situations it may play a positive role.}

Similarly, the emissivity $\varepsilon(y)$ is an even function. According to the assumption that it is proportional to the square of the field component transverse to the line of sight, we obtain $\varepsilon_1$ and $\varepsilon_2$ in the central and the peripheral areas correspondingly:
\begin{equation}
\varepsilon_1 =\kappa B_z^2\sin^2\theta \;\;\;\text{ and }\;\;\;
\varepsilon_2 =\kappa B_{\varphi}^2 (1-x^2\sin^2\theta/\rho_2^2).
\end{equation}
Let's calculate the auxiliary function $R(y)$ for three regions:
\begin{equation*}
\begin{aligned}
& R(y)=Kn B_{\varphi}\cos \varphi_0\left(l_2+y\right),\;y\in(-l_2,-l_1);\\
& R(y)=Kn B_{\varphi}\cos\varphi_0\left(l_2-l_1\right) +Kn B_z{\rm ctg\,}\theta\left(y+l_1\right),\;y\in(-l_1,l_1);\\
& R(y)=Kn B_{\varphi}\cos\varphi_0\left(l_2+y-2l_1\right)+2Kn B_z{\rm ctg\,}\theta l_1,\;y\in(l_1,l_2);
\end{aligned}
\end{equation*}
and with its help by the formula (\ref{E2}) calculate the degree and angle of polarization, which after integration take the form
\begin{equation*}
\Pi=\frac{1}{1+\gamma}\frac{\sin(\mathcal{R}_1\lambda^2)}{\mathcal{R}_1\lambda^2}+\frac{\gamma}{1+\gamma}\frac{\sin(\mathcal{R}_2\lambda^2)}{\mathcal{R}_2\lambda^2}\cos\left((\mathcal{R}_2+\mathcal{R}_1)\lambda^2+2\psi_0\right),
\end{equation*}
\begin{equation}\label{B3}
\text{and }\;\;\Psi=\left(\mathcal{R}_2+\mathcal{R}_1/2\right)\lambda^2,
\end{equation}
where for convenience we use the notation for the relative emissivity
\begin{equation}
\gamma=\frac{\varepsilon_2 (l_2-l_1)}{\varepsilon_1 l_1}=\frac{B_{\varphi}^2}{B_z^2}\frac{\rho_2^2-x^2\sin^2\theta}{\rho_2^2\sin^2\theta}\frac{l_2-l_1}{l_1}
\end{equation}
and intrinsic Faraday measures:
\begin{equation}
\mathcal{R}_1=2 Kn B_z {\rm ctg\,}\theta l_1 \;\;\text{ and }\;\;\mathcal{R}_2= Kn B_{\varphi}\, x (l_2-l_1)/\rho_2.    
\end{equation}
Note, that in the case of the line of sight perpendicular to the jet $\theta=90^\circ$ -- hereinafter we call this case the degenerate case -- we obtain $\mathcal{R}_1=0$, field $B_z$ in the central region does not rotate the polarization plane, and the main role is played by the depolarization in the peripheral region $\mathcal{R}_2\ne0$. The Faraday rotation measure $RM=\mathcal{R}_2$ in this case grows with increasing aiming distance, and the polarisation modulus is the difference between the constant and the sinc:
\begin{equation*}
\Pi=\frac{1}{1+\gamma}-\frac{\gamma}{1+\gamma}\frac{\sin(2\mathcal{R}_2\lambda^2)}{2\mathcal{R}_2\lambda^2}.
\end{equation*}
In other words at short wavelengths the radiation is partially depolarized $\Pi(0)=|1-\gamma|/|1+\gamma|$, just as it is at longer wavelengths, where $\Pi(\infty)=1/|1+\gamma|$. The measures of these depolarizations are determined by the relative intensity $\gamma$ of the magnetic fields, which in turn depends on the distance from the aiming distance $x$.

Another special case is that of a non-perpendicular line of sight, but directed exactly at the jet centre $x=0$. In this case, $\mathcal{R}_2=0$ and the depolarisation process is determined by the field in the central region rather than in the peripheral region. In this case the Faraday rotation measure exactly coincides with the analogous value for a flat galaxy $RM=\mathcal{R}_1/2$, but the degree of polarisation at $x=0$ is equal to the difference
\begin{equation}
\Pi=\frac{1}{1+\gamma}\frac{\sin(\mathcal{R}_1\lambda^2)}{\mathcal{R}_1\lambda^2}-\frac{\gamma}{1+\gamma}\cos\left(\mathcal{R}_1\lambda^2\right).
\end{equation}
Again we have that at short wavelengths the radiation is partially depolarized $\Pi(0)=|1-\gamma|/|1+\gamma|$, just as it is at longer wavelengths $\Pi(\infty)$ oscillates with amplitude $\gamma/|1+\gamma|$. The two considered cases are undoubtedly specific, in the general case the degree of polarization decreases at longer wavelengths. But the reduced polarization degree compared to the case of a flat galaxy is a typical feature of the two-zone jet model.

\section{Helical magnetic field model}\label{AppendixC}
Finally, let us consider a one-zone cylindrical model of a jet, $\rho<\rho_2$, with a helical magnetic field $\bm{B} = (B_{\varphi}\sin{\varphi}, -B_{\varphi}\cos{\varphi}, -B_z)$. We use the standard coordinate system shown in right panel of Figure~\ref{Fig1}, and the standard line of sight passing at the aiming distance $x$ from the jet axis. The angle $\varphi(y)$ in contrast to the two-zone jet model continuously depends on the $y$ coordinate and is defined by $\cos\varphi = x/(x^2+y^2)^{1/2}$ and $\sin{\varphi} =y/(x^2+y^2)^{1/2}$, where $y\in[-l_2,l_2]$ and $l_2=(\rho_2^2-x^2)^{1/2}$. The intrinsic polarization angle $\psi_0(y)$ is calculated as the angle between the northward direction $\bm{N} = (1,0,0)$ and the vector $\textbf B\times d\textbf r$ just as in the Appendix \ref{AppendixB}:
\begin{equation*}
\cos\psi_0(y)=\frac{B_z\sin\theta-B_{\varphi}\cos\varphi\cos\theta}{\left(\left(B_z\sin\theta-B_{\varphi}\cos\varphi\cos \theta\right)^2+B_{\varphi}^2\sin^2\varphi\right)^{1/2}}.
\end{equation*}
It is important that $\psi_0(y)$ is again the odd function, since the $y$ and $z$ components of the vector $\textbf B\times d\textbf r$ are odd functions with respect to $y =0$. According to the calculations in the Appendix \ref{AppendixA} it is straightforward to obtain the Faraday rotation measure for the helical field:
\begin{multline}
\frac{\Psi}{\lambda^2}=\phi(l_2)=Kn \int_{0}^{l_2}\left(B_{\varphi}\cos\varphi\sin\theta+B_z\cos\theta\right)\frac{dy}{\sin\theta}=\\=
Kn B_{\varphi} x\ln\left|{(\rho_2+l_2)}/{x}\right|+Kn B_z{\rm ctg}\theta l_2=\mathcal{R}_2+\mathcal{R}_1/2,
\end{multline}
here $B_{\varphi}={\rm const}$ and $B_{z}={\rm const}$, and the intrinsic measures defined through these fields as $\mathcal{R}_1=2Kn B_z{\rm ctg}\theta l_2$ and $\mathcal{R}_2=Kn B_{\varphi} x\ln|(\rho_2+l_2)/x|$. The first measure $\mathcal{R}_1$ that depends on the longitudinal field component has the form exactly the same as for the flat galaxy or for the two-zone model, except for the width of the active region $l_2$. The second measure $\mathcal{R}_2$ is similar to that obtained for two-zone model, see Appendix \ref{AppendixB}, but has a slightly different appearance. This is not surprising since the helical model takes into account the rotation of the azimuthal field $B_{\varphi}$ along the line of sight.

Note that, in the general case, both measures $\mathcal{R}_1$ and $\mathcal{R}_2$ depend on the profiles of the azimuthal and longitudinal components of the magnetic field in the jet. The above formulas are calculated for a constant magnetic field, but they can be similarly calculated for the increasing, e.g. $B_{\varphi}(\rho) =B_{\varphi}\rho/\rho_2$, or decreasing, e.g. $B_{\varphi}(\rho) =B_{\varphi}\rho_2/\rho$, fields correspondingly:
\begin{equation}\label{C2}
\mathcal{R}_2=Kn B_{\varphi}\,x {l_2}/{\rho_2} \;\;\;\text{ or }\;\;\;
\mathcal{R}_2=Kn B_{\varphi} \rho_2 {\rm arctg}\left({l_2}/{x}\right).
\end{equation}
In our opinion, it is not so much the specific form of each of the obtained distributions is fundamental here, but the explicit dependence of the Faraday rotation measure on the profiles of the longitudinal magnetic field $B_{z}$ -- responsible for the $x$-symmetric part of $RM$, and of the azimuthal magnetic field $B_{\varphi}$ -- responsible for the $x$-antisymmetric part of $RM$. These two dependencies allow us to constrain the inverse problem of reconstruction of the magnetic field component profiles from the dependency of the Faraday rotation measure on the aiming distance $x$. 

Similarly using the results obtained in (\ref{A6}),  we can rewrite the polarization degree for the helical model as equal to
%\begin{equation*}
%{\int\limits_{0}^{l_2}\left((b_1^2-b_2^2) \cos(2\phi\lambda^2)-2b_1b_2\sin(2\phi\lambda^2) \right)dy}\left|{\int\limits_{0}^{l_2} \left(b_1^2+b_2^2\right)dy}\right|^{-1}
%\end{equation*}
\begin{equation}
\Pi(\lambda^2)  = \frac{\int_{0}^{l_2}\left( 
(b_1^2-b_2^2) \cos(2\phi\lambda^2)-2b_1b_2\sin(2\phi\lambda^2) \right)dy}{\int_{0}^{l_2} \left(b_1^2+b_2^2\right)dy}
\end{equation}
with the emissivity $\varepsilon(y)$ proportional to the square of the perpendicular component of the magnetic field:
\begin{multline}
\varepsilon(y)=\kappa\left(b_1^2(y)+b_2^2(y)\right)\quad\text{ where }\quad b_2(y)=B_{\varphi}\sin\varphi\\
\text{and }\quad b_1(y)=B_z\sin\theta-B_{\varphi}\cos\varphi\cos\theta.
\end{multline}
It is clear that, similar to the two-zone jet model, the observed radiation is partially depolarized at short wavelengths:
\begin{equation}
\Pi(0)=\left|\frac{1-\gamma}{1+\gamma}\right|,\;\text{where}\;\gamma={\int_{0}^{l_2} b_2^2(y)\,dy}\left|{\int_{0}^{l_2}  b_1^2(y)\,dy}\right|^{-1}.
\end{equation}
For any given magnetic field component profile it is possible to calculate $\gamma$ and estimate this polarization drop, e.g. for the constant magnetic field $B_{\varphi}={\rm const}$ and $B_{z}={\rm const}$ we get $\gamma$ equal to
\begin{equation*}
\frac{B_{\varphi}^2(l_2-x\,{\rm arctg}(l_2/x))}{B_z^2\sin^2\theta l_2-x B_zB_{\varphi}\sin2\theta \ln|(\rho_2+l_2)/x|+x B_{\varphi}^2\cos^2\theta {\rm arctg}(l_2/x)}.
\end{equation*}
or for the magnetic field increasing with distance from the jet's centre $B_{\varphi}(\rho)=B_{\varphi}\rho/\rho_2$ and $B_{z}={\rm const}$ we obtain
\begin{equation}\label{C4}
\gamma=\frac{l_2^2B_{\varphi}^2}{3\rho_2^2(B_z\sin\theta-x B_{\varphi}\cos\theta /\rho_2)^2}.
\end{equation}
But what is interesting here is not the particular value of $\gamma$, but the fact of polarization depression itself.

Using the obtained formula (\ref{A6}), the degree of polarisation $\Pi$ can also be calculated. Of course, it will depend on the profile of the magnetic field, and in the simplest form of increasing to the jet edges field $B_{\varphi}(\rho)=B_{\varphi}\rho/\rho_2$ and $B_{z}={\rm const}$, when the Faraday depth $\phi$ and $y$ coordinate are proportional, the degree of polarisation will have the following form: 
\begin{equation}\label{C5}
\Pi=\frac{1}{1+\gamma}\left(1+\sqrt{3\gamma}\frac{\partial}{\partial \xi}\right)^2\frac{\sin\xi}{\xi},\;\text{ where }\;\xi=\mathcal{R}\lambda^2,
\end{equation}
the intrinsic Faraday measure, see the first formula in (\ref{C2}):
\begin{equation}
RM=\mathcal{R}/2=\mathcal{R}_2+\mathcal{R}_1/2= Kn B_{\varphi} x l_2/\rho_2+Kn B_z{\rm ctg}\,\theta l_2,    
\end{equation}
and $\gamma$ are defined by (\ref{C4}). For the considered case, the polarization degree is a combination of sincs and cosines - similar to the two-zone model. However, in our opinion, the most interesting is not the special form of the polarization degree for a particular case, but the fundamental possibility of setting the inverse problem of restoring the structure of the magnetic field by the observed polarization.

%%%%%%%%%%%%%%%%%%%%%%%%%%%%%%%%%%%%%%%%%%%%%%%%%%

% Don't change these lines
\bsp	% typesetting comment
\label{lastpage}
\end{document}